\documentclass[twocolumn,eqsecnum,prb]{revtex4}
\usepackage[T1]{fontenc}
\usepackage{amsmath,amsbsy,amssymb,graphicx}
\usepackage{times}

\let\mathbf=\boldsymbol
\def\beginABC{\begin{subequations}}
\def\endABC{\end{subequations}}

\begin{document}

\title{{\Large Spin Filter, Spin Amplifier and Other Spintronic Applications}%
\\
{\Large \ in Graphene Nanodisks }}
\author{Motohiko Ezawa}
\affiliation{Department of Applied Physics, University of Tokyo, Hongo 7-3-1, 113-8656,
Japan }

\begin{abstract}
Graphene nanodisk is a graphene derivative with a closed edge. The trigonal
zigzag nanodisk with size $N$ has $N$-fold degenerated zero-energy states. A
nanodisk can be interpletted as a quantum dot with an internal degree of
freedom. The grand state of nanodisk has been argued to be a
quasi-ferromagnet, which is a ferromagnetic-like states with a finite but
very long life time. We investigate the spin-filter effects in the system
made of nanodisks and leads based on the master equation. The finite-size
effect on spin filter is intriguing due to a reaction from the polarization
of incoming current to a quasi-ferromagnet. Analyzing the relaxation process
with the use of the Landau-Lifshitz-Gilbert equation, we explore the
response to four types of incoming currents, namely, unpolarized current,
perfectly polarized current, partially polarized current and pulse polarized
current. We propose some applications for spintronics, such as spin memory,
spin amplifier, spin valve, spin-field-effect transistor and spin diode.
\end{abstract}

\maketitle


\address{{\normalsize Department of Applied Physics, University of Tokyo, Hongo
7-3-1, 113-8656, Japan }}

\section{Introduction}

The study of spin-dependent transport phenomena has recently attracted much
attention.\cite{Murakami,OhnoInject} It has opened the way for the field of
spintronics,\cite{NittaReview,Zutic,Wolf,Prinz,OhnoIEEE} literally spin
electronics, where new device functionalities exploit both the charge and
spin degrees of freedom. There are various approaches in this sphere. For
instance, the use of a quantum-dot setup\cite{Recher,Folk} has been
proposed, which can be operated either as a spin filter to produce
spin-polarized currents or as a device to detect and manipulate spin states.
Spintronics in graphene has also been proposed\cite{Tombros,Karpan,BreyValve}
recently. In the following we address this issue in a new type of materials,
graphene nanodisks.\cite{EzawaDisk,EzawaPhysica,Fernandez,Hod,EzawaCoulomb}

Graphene nanostructure\cite{GraphExA,GraphExB,GraphExC} such as graphene
nanoribbons\cite%
{Fujita,EzawaPRB,Brey,Rojas,Son,Barone,Kim,Avouris,Xu,Ozyilmaz} and graphene
nanodisks\cite{EzawaDisk,EzawaPhysica,Fernandez,Hod,EzawaCoulomb} could be
basic components of future nanoelectronic and spintronic devices. Graphene
nanodisks are nanometer-scale disk-like materials which have closed edges.
They are constructed by connecting several benzenes, some of which have
already been manufactured by soft-landing mass spectrometry.\cite%
{Rader,Kim,Berger} There are varieties of nanodisks, among which zigzag
trigonal nanodisks with size $N$ are prominent because they have $N$-fold
degenerated zero-energy states. We have already shown\cite{EzawaDisk} that
spins make a ferromagnetic order and that the relaxation time is finite but
quite large even if the size $N$ is very small. We refer to this property as
quasi-ferromagnet. Furthermore we have argued\cite{EzawaCoulomb} that a
nanodisk behaves as if it were a quantum dot with an internal degree of
freedom, where the conductance exhibits a peculiar series of Coulomb
blockade peaks.

In this paper we make an investigation of the spin current in the
zero-energy sector of the trigonal zigzag nanodisk. We first analyze how the
spin of a nanodisk filters the spin of the current by assuming that the
nanodisk is a rigid ferromagnet. The fact that the direct and exchange
Coulomb interactions are of the same of magnitude plays an important role.
However the nanodisk is not a rigid ferromagnet but a quasi-ferromagnet.
Hence an intriguing reaction phenomenon occurs: Namely, the spin of the
nanodisk can be controlled by the spin of the current. Using these
properties we propose some applications for spintronic devices, such as spin
memory, spin amplifier, spin valve, spin-field-effect transistor and spin
diode.

This paper is organized as follows. In Sec.\ref{SecNanodisk}, we summarize
the basic notion of the nanodisk. The low-energy physics is described by
electrons in the $N$-fold degenerated zero-energy sector, which form a
quasi-ferromagnet due to a large exchange interaction. In Sec.\ref%
{SecSpinFilter}, we investigate the spin-filter effects based on the master
equation. In Sec.\ref{SecReaction}, we analyze the reaction to the spin of
the nanodisk from the spin of electrons in the current. We discuss the
relaxation process using the Landau-Lifshitz-Gilbert equation. In Sec.\ref%
{SecApplication}, we propose various spintronic devices and explore their
properties.

\section{Nanodisk Quasi-Ferromagnets}

\label{SecNanodisk}

\subsection{Zero-Energy Sector}

Graphene nanodisks are graphene derivatives which have closed edges. The
Hamiltonian is defined by%
\begin{equation}
H=\sum_{i}\varepsilon _{i}c_{i}^{\dagger }c_{i}+\sum_{\left\langle
i,j\right\rangle }t_{ij}c_{i}^{\dagger }c_{j},  \label{HamilTB}
\end{equation}%
where $\varepsilon _{i}$ is the site energy, $t_{ij}$ is the transfer
energy, and $c_{i}^{\dagger }$ is the creation operator of the $\pi $
electron at the site $i$. The summation is taken over all nearest
neighboring sites $\left\langle i,j\right\rangle $. Owing to their
homogeneous geometrical configuration, we may take constant values for these
energies, $\varepsilon _{i}=\varepsilon _{\text{F}}$ and $t_{ij}=t\approx
2.70$eV.

We have searched for zero-energy states or equivalently metallic states in
graphene nanodisks with various sizes and shapes.\cite{EzawaDisk} The
emergence of zero-energy states is surprisingly rare. Among typical
nanodisks, only trigonal zigzag nanodisks have degenerate zero-energy states
and show metallic ferromagnetism, where the degeneracy can be controlled
arbitrarily by designing the size.

\begin{figure}[h]
\includegraphics[width=0.46\textwidth]{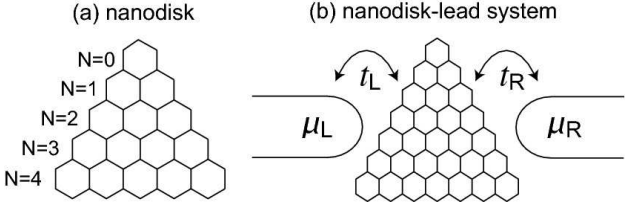}
\caption{(a) Trigonal zigzag nanodisks. The size parameter $N$ is defined in
this way. The number of carbon atoms is given by $N_{\text{C}}=N^{2}+6N+6$.
(b) The nanodisk-lead system. The nanodisk with $N=7$ is connected to the
right and left leads by tunneling coupling $t_{\text{R}}$ and $t_{\text{L}}$%
. The chemical potential is $\protect\mu _{\text{R}}$ or $\protect\mu _{%
\text{L}}$ at each lead. }
\label{FigNanodisk}
\end{figure}

Trigonal zigzag nanodisks are specified by the size parameter $N$ as in Fig.%
\ref{FigNanodisk}(a). The size-$N$ nanodisk has $N$-fold degenerated
zero-energy states,\cite{EzawaDisk} where the gap energy is as large as a
few eV. Hence it is a good approximation to investigate the
electron-electron interaction physics only in the zero-energy sector, by
projecting the system to the subspace made of those zero-energy states. The
zero-energy sector consists of $N$ orthonormal states $|f_{\alpha }\rangle $%
, $\alpha =1,2,\cdots ,N$, together with the SU(N) symmetry. We can expand
the wave function of the state $|f_{\alpha }\rangle $,%
\begin{equation}
f_{\alpha }(\boldsymbol{x})=\sum_{i}\omega _{i}^{\alpha }\varphi _{i}(%
\boldsymbol{x}),  \label{EqA}
\end{equation}%
in terms of the Wannier function $\varphi _{i}(\boldsymbol{x})$ localized at
the site $i$. The amplitude $\omega _{i}^{\alpha }$ is calculable by
diagonalizing the Hamiltonian (\ref{HamilTB}).

\subsection{Quasi-Ferromagnets}

We include the Coulomb interaction between electrons in the zero-energy
sector.\cite{EzawaCoulomb} It is straightforward to rewrite the Coulomb
Hamiltonian as%
\begin{align}
H_{\text{D}}^{\text{eff}}=& \sum_{\alpha \geq \beta }U_{\alpha \beta
}n\left( \alpha \right) n\left( \beta \right)  \notag \\
& -\frac{1}{2}\sum_{\alpha >\beta }J_{\alpha \beta }[4\mathbf{S}(\alpha
)\cdot \mathbf{S}(\beta )+n\left( \alpha \right) n\left( \beta \right) ],
\label{HamilFerro}
\end{align}%
where $U_{\alpha \beta }$ and $J_{\alpha \beta }$ are the Coulomb energy and
the exchange energy between electrons in the states $|f_{\alpha }\rangle $
and $|f_{\beta }\rangle $. Here, $n\left( \alpha \right) $ is the number
operator and $\mathbf{S}(\alpha )$ is the spin operator,\beginABC%
\begin{align}
n\left( \alpha \right) & =d_{\sigma }^{\dag }(\alpha )d_{\sigma }(\alpha ),
\\
\mathbf{S}(\alpha )& =\frac{1}{2}d_{\sigma }^{\dag }(\alpha )\mathbf{\tau }%
_{\sigma \sigma ^{\prime }}d_{\sigma ^{\prime }}(\alpha ),
\end{align}%
\endABC with $d_{\sigma }(\alpha )$ the annihilation operator of electron
with spin $\sigma =\uparrow ,\downarrow $ in the state $|f_{\alpha }\rangle $%
: $\mathbf{\tau }$ is the Pauli matrix.

The remarkable feature is that there exists a large overlap between the wave
functions $f_{\alpha }(\boldsymbol{x})$ and $f_{\beta }(\boldsymbol{x})$, $%
\alpha \neq \beta $, since the state $|f_{\alpha }\rangle $ is an ensemble
of sites as in (\ref{EqA}) and identical sites are included in $|f_{\alpha
}\rangle $ and $|f_{\beta }\rangle $. Consequently, the dominant
contributions come from the on-site Coulomb terms not only for the Coulomb
energy but also for the exchange energy: Indeed, it follows that $U_{\alpha
\beta }=J_{\alpha \beta }$ in the on-site approximation. We thus obtain%
\begin{equation}
U_{\alpha \beta }\simeq J_{\alpha \beta }\simeq {U}\sum_{i}(\omega
_{i}^{\alpha }\omega _{i}^{\beta })^{2},  \label{EqY}
\end{equation}%
where%
\begin{equation}
U\equiv \sum_{i,j}\int \!d^{3}xd^{3}y\;\varphi _{i}^{\ast }(\boldsymbol{x}%
)\varphi _{i}(\boldsymbol{x})V(\mathbf{x}-\mathbf{y})\varphi _{j}^{\ast }(%
\boldsymbol{y})\varphi _{j}(\boldsymbol{y})
\end{equation}%
with the Coulomb potential $V(\mathbf{x}-\mathbf{y})$. The Coulomb energy $U$
is of the order of $1$eV because the lattice spacing of the carbon atoms is $%
\sim 1$\r{A}\ in graphene.

Since the exchange energy $J_{\alpha \beta }$ is as large as the Coulomb
energy $U_{\alpha \beta }$, the spin stiffness $J_{\alpha \beta }$ is quite
large. Furthermore, we have checked\cite{EzawaCoulomb} numerically that all $%
J_{\alpha \beta }$ are of the same order of magnitude for any pair of $%
\alpha $ and $\beta $, implying that the SU(N) symmetry is broken but not so
strongly in the Hamiltonian (\ref{HamilFerro}). It is a good approximation
to start with the exact SU($N$) symmetry. Then, the zero-energy sector is
described by the SU(N) Heisenberg-Hubbard model,%
\begin{align}
H_{\text{D}}=& -J\sum_{\alpha \neq \beta }\mathbf{S}\left( \alpha \right)
\cdot \mathbf{S}\left( \beta \right)  \notag \\
& +\left( \frac{U}{2}-\frac{J}{4}\right) \sum_{\alpha \neq \beta }n\left(
\alpha \right) n\left( \beta \right) +U\sum_{\alpha }n\left( \alpha \right) ,
\label{HamilD}
\end{align}%
with $J\approx U$. We rewrite $H_{\text{D}}$ as 
\begin{align}
H_{\text{D}}=& -J\left[ S_{\text{tot}}^{2}-\sum_{\alpha }\mathbf{S}%
^{2}\left( \alpha \right) \right] +\left( \frac{U}{2}-\frac{J}{4}\right) n_{%
\text{tot}}^{2}  \notag \\
& +\left( \frac{U}{2}+\frac{J}{4}\right) \sum_{\alpha }n\left( \alpha \right)
\notag \\
=& -JS_{\text{tot}}^{2}+\left( \frac{U}{2}-\frac{J}{4}\right) n_{\text{tot}%
}^{2}+\left( \frac{U}{2}+J\right) n_{\text{tot}}  \label{HamilDx}
\end{align}%
in terms of the total spin,%
\begin{equation}
\mathbf{S}_{\text{tot}}=\sum_{\alpha }\mathbf{S}\left( \alpha \right) ,
\end{equation}%
and the total electron number,%
\begin{equation}
n_{\text{tot}}=\sum_{\alpha }n\left( \alpha \right) .
\end{equation}%
At the half-filling, the eigenstate of the Hamiltonian $H_{\text{D}}$ is
labeled as $\left\vert s\right\rangle =\left\vert n_{\text{tot}%
},l,m\right\rangle $, where $l$ is the total spin-angular momentum and $m$
is its z-component. Based on this Hamiltonian we have shown\cite{EzawaDisk}
that the relaxation time of the ferromagnetic-like spin polarization is
quite large even if the size of trigonal zigzag nanodisks is quite small.
Such a nanodisk may be called a quasi-ferromagnet. We refer to the total
spin $\mathbf{S}_{\text{tot}}$ of a nanodisk as the nanodisk spin.

\subsection{Nanodisk-Lead System}

We consider a system made of a nanodisk connected by two metallic leads [Fig.%
\ref{FigNanodisk}(b)]. The model Hamiltonian is given by%
\begin{equation}
H=H_{\text{L}}+H_{\text{TL}}+H_{\text{TR}}+H_{\text{D}},  \label{TotalHamil}
\end{equation}%
where $H_{\text{D}}$ is the Hamiltonian (\ref{HamilD}) of a nanodisk, and $%
H_{\text{L}}$ is the lead Hamiltonian 
\begin{equation}
H_{\text{L}}=\sum_{k\sigma }\varepsilon \left( k\right) \left( c_{k\sigma }^{%
\text{L}\dagger }c_{k\sigma }^{\text{L}}+c_{k\sigma }^{\text{R}\dagger
}c_{k\sigma }^{\text{R}}\right) .
\end{equation}%
The Hamiltonian $H_{\text{L}}$ describes a noninteracting electron gas in
the leads with $\varepsilon \left( k\right) =\hbar ^{2}\mathbf{k}^{2}/2m$,
while $H_{\text{TL}}$ and $H_{\text{TR}}$ are the transfer Hamiltonian
between the left (L) and right (R) leads and the nanodisk, respectively,%
\beginABC\label{HamilT}%
\begin{align}
H_{\text{TL}}=& t_{\text{L}}\sum_{k\sigma }\sum_{\alpha }\left( c_{k\sigma
}^{\text{L}\dagger }d_{\sigma }(\alpha )+d_{\sigma }^{\dagger }(\alpha
)c_{k\sigma }^{\text{L}}\right) , \\
H_{\text{TR}}=& t_{\text{R}}\sum_{k\sigma }\sum_{\alpha }\left( c_{k\sigma
}^{\text{R}\dagger }d_{\sigma }(\alpha )+d_{\sigma }^{\dagger }(\alpha
)c_{k\sigma }^{\text{R}}\right) ,
\end{align}%
\endABC with $t_{\chi }$ the tunneling coupling constant: We have assumed
that the spin does not flip in the tunneling process.

The nanodisk-lead system looks similar to that of the $N$-dot system.\cite%
{TaruchaRev} However, there exists a crucial difference. On one hand, in the
ordinary $N$-dot system, an electron hops from one dot to another dot. On
the other hand, in our nanodisk system, the index $\alpha $ of the
Hamiltonian runs over the $N$-fold degenerate states and not over the sites.
According to the Hamiltonian (\ref{HamilT}), an electron does not hop from
one state to another state. Hence, it is more appropriate to regard our
nanodisk as a one-dot system with an internal degree of freedom.

\section{Spin Filter}

\label{SecSpinFilter}

The aim of this paper is to investigate a nanodisk as a spin filter\cite%
{NittaPRL02} based on the Hamiltonian (\ref{TotalHamil}). The setup we
consider is a lead-nanodisk-lead system [Fig.\ref{FigNanodisk}(b)], where an
electron makes a tunnelling from the left lead to the nanodisk and then to
the right lead. We investigate how the electron spin is affected by the
nanodisk spin during the transport process.

The dynamics of the nanodisk system is described by the master equation, 
\begin{equation}
\frac{\partial \rho \left( s\right) }{\partial t}=\sum_{s^{\prime }}\left[
W\left( s,s^{\prime }\right) \rho \left( s^{\prime }\right) -W\left(
s^{\prime },s\right) \rho \left( s\right) \right] ,
\end{equation}%
where $\rho \left( s\right) $ represents the probability to find the system
in the state $\left\vert s\right\rangle =\left\vert n_{\text{tot}%
},l,m\right\rangle $, and $W\left( s^{\prime },s\right) $ is the transition
rate between the states $\left\vert s\right\rangle $ and $\left\vert
s^{\prime }\right\rangle $. The master equation describes a stochastic
evolution in the space spanned by the states $\left\vert s\right\rangle $.
Jumps between different states are assumed to be Markovian. The stationary
solution is given by the detailed balance condition $\partial \rho /\partial
t=0$.

Since the Hamiltonian $H_{\text{D}}$ only depends on $S_{\text{tot}}^{2}$,
the probability $\rho (s)$ has no dependence on the spin component $m$. Then
it is convenient to do a sum over $m$, 
\begin{equation}
\overline{\rho }\left( \overline{s}\right) =\sum_{m}\rho \left( s\right)
\end{equation}%
with $\left\vert \overline{s}\right\rangle =\left\vert n_{\text{tot}%
},l\right\rangle $. The master equation is rewritten as%
\begin{equation}
\frac{\partial \overline{\rho }\left( \overline{s}\right) }{\partial t}%
=\sum_{s^{\prime }}\left[ W\left( \overline{s},\overline{s}^{\prime }\right) 
\overline{\rho }\left( \overline{s}^{\prime }\right) -W\left( \overline{s}%
^{\prime },\overline{s}\right) \overline{\rho }\left( \overline{s}\right) %
\right] .
\end{equation}%
In the following we denote $\overline{\rho }\left( \overline{s}\right) $ as $%
\rho \left( s\right) $ for simplicity.

In this paper we consider the small coupling limit, $t_{\chi }\ll J$, $\chi
= $L, R, where the dominant process is the sequential tunneling: It is of
the order of $\left\vert t_{\chi }\right\vert ^{2}$, while the cotunneling
process of the order of $\left\vert t_{\chi }\right\vert ^{4}$.

In the sequential-tunneling regime the transition rate, $W=W_{\text{L}}+W_{%
\text{R}}$, is obtained by the Fermi's golden rule,\cite%
{Beenakker,Recher,Golovach}%
\begin{align}
W_{\chi }\left( s^{\prime },s\right) =& \Gamma _{\chi }f_{\chi }\left(
\Delta _{s^{\prime }s}\right) \delta _{n^{\prime },n+1}  \notag \\
& +\Gamma _{\chi }\left[ 1-f_{\chi }\left( \Delta _{s^{\prime }s}\right) %
\right] \delta _{n^{\prime },n-1}.  \label{TransRate}
\end{align}%
Let us explain notations. First,%
\begin{equation}
f_{\chi }\left( \varepsilon \right) =\left\{ 1+\exp \left[ \left(
\varepsilon -\mu _{\chi }\right) /k_{\text{B}}T\right] \right\} ^{-1}
\end{equation}%
is the Fermi function at temperature $T$, with $\mu _{\chi }$ the chemical
potential at the right or left lead, $\chi =$R,L. We set 
\begin{equation}
\mu _{\text{L}}=\mu +\Delta \mu ,\quad \mu _{\text{R}}=\mu -\Delta \mu .
\end{equation}%
Second, $\Delta _{s^{\prime }s}$ is the energy difference between the two
states $|s^{\prime }\rangle $ and $|s\rangle $,%
\begin{equation}
\Delta _{s^{\prime }s}=E_{s^{\prime }}-E_{s}.  \label{Gap}
\end{equation}%
Third,%
\begin{equation}
\Gamma _{\chi }=2\pi \nu \left\vert A_{\chi ,ss^{\prime }}^{\sigma
}\right\vert ^{2}  \label{TunneRate}
\end{equation}%
is the tunneling rate through the right or left lead, where $\nu
=\sum_{k}\delta \left( \varepsilon _{\text{F}}-\varepsilon _{k}\right) $ is
the density of states at the Fermi level $\varepsilon _{\text{F}}$ in the
leads, which is a constant, and 
\begin{equation}
A_{\chi ,ss^{\prime }}^{\sigma }=t_{\chi }\sum_{\alpha }\left\langle
s^{\prime }\right\vert d_{\alpha \sigma }\left\vert s\right\rangle .
\end{equation}%
It follows that the tunneling rate $\Gamma _{\chi }$ is a constant for the
states $|s\rangle $ and $|s^{\prime }\rangle $ connected by a Markov step,
and zero otherwise.

The current through the nanodisk can be written as 
\begin{equation}
I_{\sigma }=\sum_{s,s^{\prime }}\left[ W_{\text{R}}\left( s^{\prime
},s\right) \rho \left( s\right) -W_{\text{R}}\left( s,s^{\prime }\right)
\rho \left( s^{\prime }\right) \right] ,  \label{SpinCurre}
\end{equation}%
where $\sum_{s,s^{\prime }}$ runs over the states $|s\rangle $ and $%
|s^{\prime }\rangle $ such that $\left\langle s^{\prime }\right\vert
d_{\alpha \sigma }\left\vert s\right\rangle \neq 0$. Thus the current $%
I_{\sigma }$ depend on the spin configuration of electrons in the nanodisk.

\begin{figure}[h]
\includegraphics[width=0.4\textwidth]{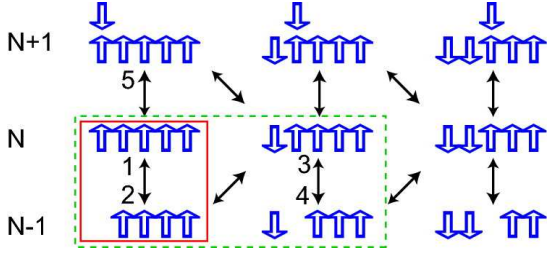}
\caption{Markov chain of spin configurations in the $N=5$ nanodisk. An arrow
($\updownarrow $) indicates each Markov step between two states with
different spin configurations. The region surrounded by red solid lines
shows the most dominant process for spin-polarized current. In this
approximation there is no down-spin polarized current. The region surrounded
by green dotted lines yields the next dominant contribution, which allows a
down-spin polarized current as well.}
\label{FigState}
\end{figure}

We have argued that the nanodisk is a quasi-ferromagnet. For the sake of
simplicity, we start with the simplification that the ground state is a
ferromagnet with polarized up-spins. When the transition interaction is
small enough, a single electron tunnels at once, namely, the electron number
in the nanodisk increases or decreases by one, $n_{\text{tot}}^{\prime }=n_{%
\text{tot}}\pm 1$. First of all, we notice that, when $n_{\text{tot}%
}^{\prime }=n_{\text{tot}}+1$, the tunneling of spin-up electrons is blocked
by the Pauli's exclusion principle. The Markov chain of spin configurations
contain only a finite number of states. For instance it is $9$ in the case
of $N=5$ nanodisk. We number them as indicated in Fig.\ref{FigState}.

We calculate the energy $E_{s}$ of various states $\left\vert s\right\rangle
=\left\vert n_{\text{tot}},l\right\rangle $ in this chain. According to the
above numbering convention (Fig.\ref{FigState}), they are%
\begin{align}
E_{1}=& E_{N,N/2},\quad E_{2}=E_{N-1,(N-1)/2},\quad E_{3}=E_{N,N/2-1}, 
\notag \\
E_{4}=& E_{N-1,(N-3)/2},\quad E_{5}=E_{N+1,(N-1)/2},\quad \cdots .
\end{align}%
Based on the effective Hamiltonian (\ref{HamilDx}) they are calculated as
follows,\beginABC%
\begin{align}
E_{1}=& \frac{U-J}{2}N^{2}+\frac{U+J}{2}N\approx UN, \\
E_{2}=& \frac{U-J}{2}\left( N-1\right) ^{2}+\frac{U+J}{2}\left( N-1\right)
\approx U(N-1), \\
E_{3}=& \frac{U-J}{2}N^{2}+\frac{U+3J}{2}N\approx 2UN, \\
E_{4}=& \frac{U-J}{2}\left( N-1\right) ^{2}+\frac{U+3J}{2}\left( N-1\right)
\approx 2U(N-1), \\
E_{5}=& \frac{U-J}{2}N^{2}+\frac{3}{2}\left( U+J\right) N-J\approx 3UN-U,
\end{align}%
\endABC and so on.

The order of the energies is%
\begin{equation}
E_{2}<E_{1}<E_{4}<E_{3}<E_{5}<\cdots .
\end{equation}%
The probability to find the state $\left\vert s\right\rangle $ contains the
Boltzman factor $\exp (-E_{s}/k_{\text{B}}T)$. Hence, the most dominant
Markov chain consists only of the states $|1\rangle $ and $|2\rangle $, and
it is denoted as $\left( 1\leftrightarrow 2\right) $. This process yields
the spin-up current. We consider the second dominant process. Though $%
E_{4}<E_{3}$, the state $|4\rangle $ can only be reached via the state $%
|3\rangle $, as illustrated in Fig.\ref{FigState}. Namely, the second
dominant chain is $\left( 1\leftrightarrow 2\leftrightarrow 3\leftrightarrow
4\right) $, which allows the spin-down current as well as the spin-up
current.

First we analyze the most dominant process, which involves only the ground
state and the first excited state (Fig.\ref{FigState}). The currents are
given by\beginABC%
\begin{align}
I_{\uparrow }=& \frac{I_{0}}{2}\left[ f_{\text{L}}\left( E_{1}-E_{2}\right)
-f_{\text{R}}\left( E_{1}-E_{2}\right) \right]  \notag \\
=& \frac{I_{0}}{2}\frac{\sinh (\Delta \mu /k_{\text{B}}T)}{\cosh (\Delta \mu
/k_{\text{B}}T)+\cosh ((U-\mu )/k_{\text{B}}T)}, \\
I_{\downarrow }=& 0,
\end{align}%
\endABC where%
\begin{equation}
I_{0}=\frac{e\Gamma _{\text{L}}\Gamma _{\text{R}}}{\Gamma _{\text{L}}+\Gamma
_{\text{R}}}
\end{equation}%
in terms of the tunneling rate $\Gamma _{\chi }$ given by (\ref{TunneRate}).

The above expression can be further simplified in the following two limits.
In the high-temperature limit ($\mu ,\Delta \mu \ll k_{\text{B}}T$) the
sequential tunneling current takes a simple form 
\begin{align}
I_{\uparrow }\simeq & \frac{I_{0}}{2}\Delta \mu \lim_{\Delta \mu \rightarrow
0}\frac{\partial f_{\text{L}}\left( E_{1}-E_{2}\right) }{\partial \Delta \mu 
}  \notag \\
=& I_{0}\frac{\Delta \mu }{8k_{\text{B}}T}\cosh ^{-2}\left[ \frac{\Delta
_{12}-\mu }{2k_{\text{B}}T}\right] ,
\end{align}%
where $\Delta _{s^{\prime }s}$ is defined in (\ref{Gap}). On the other hand,
in the low-temperature limit ($k_{\text{B}}T\ll \mu ,\Delta \mu $) the
current is given by%
\begin{equation}
I_{\uparrow }=\frac{I_{0}}{2}\left[ \theta \left( \mu _{L}-\Delta
_{12}\right) -\theta \left( \mu _{R}-\Delta _{12}\right) \right]
\end{equation}%
with the step function 
\begin{equation}
\theta \left( x\right) =\left\{ 
\begin{array}{cc}
1,\quad & x>0 \\ 
0,\quad & x<0%
\end{array}%
\right. .
\end{equation}%
The current $I_{\uparrow }$ flows in the triangle domain ($\mu _{L}>\Delta
_{12}>\mu _{R}$) in the $(\Delta \mu ,\mu )$ plain.

\begin{figure}[h]
\includegraphics[width=0.46\textwidth]{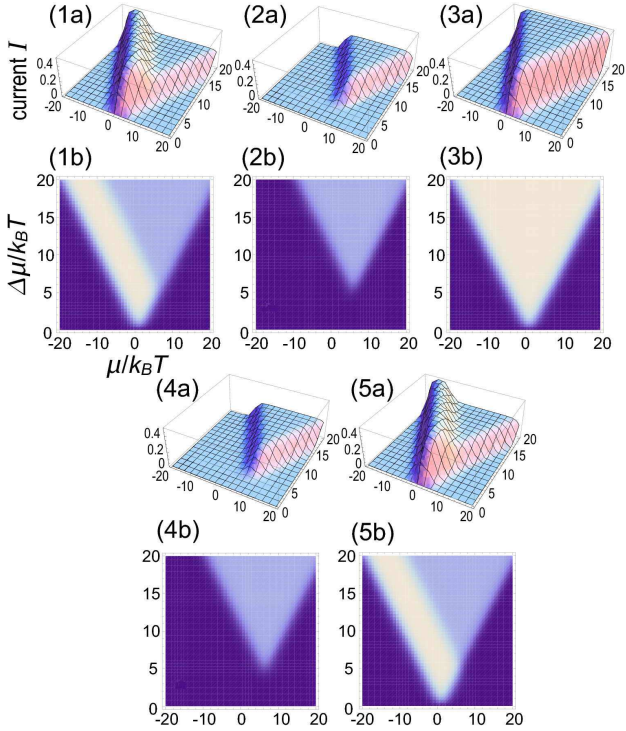}
\caption{Spin-dependent currents against $\protect\mu /k_{\text{B}}T$ and $%
\Delta \protect\mu /k_{\text{B}}T$ in the $N=10$ nanodisk. We have set $U/k_{%
\text{B}}T=1$. The horizontal $x$-axis is the chemical potential $\protect%
\mu /k_{\text{B}}T$, and the horizontal $y$-axis is the bias voltage $\Delta 
\protect\mu /k_{\text{B}}T$. (a) The vertical $z$-axis is the current. (b)
More currents flow in brighter regions. (1) The current $I_{1}$ induced by
the process $1\leftrightarrow 2$, which produces the up-spin polarized
current. (2) The current $I_{2}$ induced by the process $3\leftrightarrow 4$%
, which produces the up-spin polarized current. (3) The sum of the above two
processes; $I_{\uparrow }=I_{1}+I_{2}$. (4) The current $I_{4}$ induced by
the process $2\leftrightarrow 3$, which produces the down-spin polarized
current; $I_{\downarrow }=I_{4}$. (5) The difference $\Delta I$ between the
up-spin and down-spin current, $\Delta I=I_{\uparrow }-I_{\downarrow
}=I_{1}+I_{2}-I_{4}$.}
\label{FigSpinFilter}
\end{figure}

In order to search for the down-spin polarized current $I_{\downarrow }$, it
is necessary to analyze the second dominant process. It includes four states 
$\left( 1\leftrightarrow 2\leftrightarrow 3\leftrightarrow 4\right) $, as
shown in the Fig.\ref{FigState}. In this approximation, the transition
matrix becomes a tridiagonal matrix, and we can solve the master equations
exactly. The nonequilibrium density of states is calculated as%
\begin{equation}
\rho _{s}=\frac{\widetilde{\rho }_{s}}{\text{Tr}\widetilde{\rho }_{s}}%
,\qquad \widetilde{\rho }_{s}=\prod_{i=1}^{s-1}\frac{W\left( i+1,i\right) }{%
W\left( i,i+1\right) },
\end{equation}%
where the index $s=1,2,3,4$ stands for the four states with the energy $E_{s}
$. We can explicitly calculate $\widetilde{\rho }_{s}$ as%
\begin{equation}
\widetilde{\rho }_{s}=e^{-\beta \widetilde{E}_{s}}\prod_{i=1}^{s-1}\frac{%
e^{\beta \widetilde{E}_{i}}+e^{\beta \widetilde{E}_{i+1}}\cosh \Delta \mu }{%
e^{\beta \widetilde{E}_{i}}\cosh \Delta \mu +e^{\beta \widetilde{E}_{i+1}}},
\end{equation}%
with $\widetilde{E}_{s}=E_{s}-\mu n_{\text{tot}}$. The current $I_{\sigma }$
is calculable with the use of (\ref{SpinCurre}) and (\ref{TransRate}), whose
result we show in the Fig.\ref{FigSpinFilter}. In particular we have%
\begin{align}
I_{\downarrow }=& \frac{I_{0}}{4}\left[ \theta \left( \mu _{L}-\Delta
_{12}\right) -\theta \left( \mu _{R}-\Delta _{12}\right) \right]   \notag \\
& \times \left[ \theta \left( \mu _{L}-\Delta _{32}\right) -\theta \left(
\mu _{R}-\Delta _{32}\right) \right] 
\end{align}%
in the zero-temperature limit. The current $I_{\downarrow }$ flows in the
triangle domain ($\mu _{L}>\Delta _{12}>\mu _{R}$ and $\mu _{L}>\Delta
_{32}>\mu _{R}$) in the $(\Delta \mu ,\mu )$ plain.

\section{Finite-size effects: Reaction to quasi-ferromagnet}

\label{SecReaction}

In the conventional spin filter the ferromagnet is very rigid. On the other
hand, the life time is finite in the case of a nanodisk, though it is quite
long in spite of its small size. We expect a reaction to the nanodisk spin
from the spin-polarized current, provided the nanodisk spin is not
controlled externally, say, by magnetic field. This effect is very
interesting, since it is an intrinsic nature to quasi-ferromagnets.

We inject an electron into a nanodisk. Let $\Delta _{\uparrow \uparrow }$ or 
$\Delta _{\uparrow \downarrow }$ be the energy increase when the spin of an
injected electron is parallel or anti-parallel to the nanodisk spin.
According to the Hamiltonian (\ref{HamilD}) they are 
\begin{align}
\Delta _{\uparrow \uparrow }=\Delta _{21}=& -U,  \notag \\
\Delta _{\uparrow \downarrow }=\Delta _{31}=& UN.
\end{align}%
When the direction between the nanodisk spin and the electron spin is $%
\theta $, the energy increase is given by%
\begin{equation}
\langle H_{\text{D}}^{\text{eff}}\rangle =\Delta _{\uparrow \uparrow }\cos 
\frac{\theta }{2}\langle c_{\uparrow }^{\dagger }c_{\uparrow }\rangle
+\Delta _{\uparrow \downarrow }\sin \frac{\theta }{2}\langle c_{\downarrow
}^{\dagger }c_{\downarrow }\rangle ,
\end{equation}%
where $\left\langle c_{\sigma }^{\dagger }c_{\sigma }\right\rangle $
represents the probability of finding the injected electron to have spin $%
\sigma $.

Based on the spin-rotational symmetry we may write the effective Hamiltonian
as%
\begin{equation}
H_{\text{D}}^{\text{eff}}=-J_{\text{sd}}\mathbf{S}\cdot c^{\dagger }\mathbf{%
\sigma }c=-M\mathbf{n}\cdot c^{\dagger }\mathbf{\sigma }c,
\end{equation}%
where 
\begin{equation}
J_{\text{sd}}=MS_{\text{tot}}=\Delta _{\uparrow \downarrow }-\Delta
_{\uparrow \uparrow }=U\left( N+1\right) ,
\end{equation}
and $\mathbf{n}=\mathbf{S}_{\text{tot}}/S_{\text{tot}}$ is the normalized
spin with $S_{\text{tot}}=|\mathbf{S}_{\text{tot}}|$.

We introduce the Gilbert damping term phenomenologically,%
\begin{equation}
H_{\alpha }=\alpha \mathbf{\dot{n}}^{2},
\end{equation}%
where $\alpha $ is a dimensionless constant ($\alpha \approx 0.01$). Using
the variational method for the Hamiltonian $H_{\text{D}}^{\text{eff}%
}+H_{\alpha }$, we obtain the Landau-Lifshitz-Gilbert equation,%
\begin{equation}
\frac{\partial \mathbf{n}}{\partial t}=\gamma \mathbf{B}_{\text{eff}}\times 
\mathbf{n}-\alpha \mathbf{n}\times \frac{\partial \mathbf{n}}{\partial t},
\label{LLG}
\end{equation}%
where $\mathbf{B}_{\text{eff}}$ is the effective magnetic field produced by
the injected electron spin,%
\begin{equation}
\gamma \mathbf{B}_{\text{eff}}=\frac{1}{\hbar S_{\text{tot}}}\left\langle 
\frac{\partial H_{\text{D}}^{\text{eff}}}{\partial \mathbf{n}}\right\rangle
=-\frac{M}{\hbar S_{\text{tot}}}\left\langle c^{\dagger }\mathbf{\sigma }%
c\right\rangle ,  \label{EffecB}
\end{equation}%
with the gyromagnetic ratio $\gamma $.

For definiteness we now inject the up-spin electron to the nanodisk. The
effective magnetic field is%
\begin{equation}
\gamma \mathbf{B}_{\text{eff}}=(0,0,\gamma B_{\text{eff}})=-\frac{M}{\hbar
S_{\text{tot}}}\left\langle c^{\dagger }\mathbf{\sigma }c\right\rangle .
\end{equation}%
We investigate the dynamics of the normalized spin $\mathbf{n}$ of the
nanodisk under this field. In the polar coordinate, setting 
\begin{equation}
\mathbf{n}=\left( \sin \theta \cos \phi ,\sin \theta \sin \phi ,\cos \theta
\right) ,
\end{equation}%
we rewrite the Landau-Lifshitz-Gilbert equation (\ref{LLG}) as%
\begin{equation}
\dot{\theta}=\alpha \dot{\phi}\sin \theta ,\qquad \dot{\phi}\sin \theta
=\gamma B_{\text{eff}}\sin \theta -\alpha \dot{\theta}.
\end{equation}%
This is equivalent to%
\begin{equation}
\dot{\theta}=\frac{\alpha \gamma B_{\text{eff}}}{1+\alpha ^{2}}\sin \theta
,\qquad \dot{\phi}=\frac{\gamma B_{\text{eff}}}{1+\alpha ^{2}},
\end{equation}%
which we can solve explicitly,%
\begin{align}
\theta \left( t\right) & =2\tan ^{-1}\exp \left[ -\left( t-t_{0}\right)
/\tau _{\text{filter}}\right] ,  \notag \\
\phi \left( t\right) & =\frac{\gamma B_{\text{eff}}}{1+\alpha ^{2}}t,
\label{angle}
\end{align}%
where $t_{0}$ is an integration constant, and $\tau _{\text{filter}}$ is the
relaxation time given by%
\begin{equation}
\tau _{\text{filter}}=\frac{1+\alpha ^{2}}{2\alpha \gamma \left\vert B_{%
\text{eff}}\right\vert }\propto N.  \label{RelaxTime}
\end{equation}%
It is proportional to $S_{\text{tot}}/M$ and hence proportional to the disk
size $N$ because of (\ref{EffecB}). The initial phase $\theta _{0}$ is
related to the parameter $t_{0}$ as%
\begin{equation}
\theta \left( 0\right) =2\tan ^{-1}\exp \left[ t_{0}/\tau _{\text{filter}}%
\right] .  \label{InitialPhase}
\end{equation}%
Thus, $t_{0}=0$ corresponds to 
\begin{equation}
\theta \left( 0\right) =\frac{\pi }{2}.
\end{equation}%
We use the parameter $t_{0}$ instead of the initial phase $\theta \left(
0\right) $ for simplicity.

\begin{figure}[h]
\includegraphics[width=0.33\textwidth]{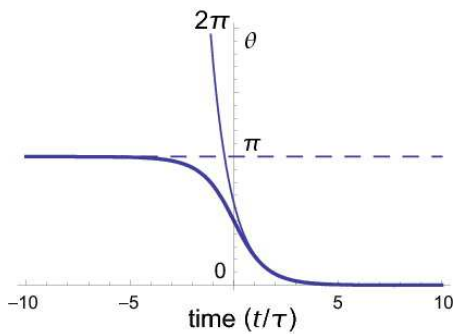}
\caption{The bold curve is the relaxation process of spin in nanodisk. The
horizontal axis is the time $t/\protect\tau $ and the vertical axis is the
angle $\protect\theta $. The thin curve is the asymptotic function $\protect%
\theta =2\exp \left[ -\left( t-t_{0}\right) /\protect\tau _{\text{filter}}%
\right] $ of the bold curve for $t\gg t_{0}$. The dotted line is the
asymptotic value for $t\ll t_{0}$.}
\label{LLGRelax}
\end{figure}

The time scale for the direction of the nanodisk spin to align with that of
the spin-polarized current is $\tau _{\text{filter}}$, where the effective
magnetic field (\ref{EffecB}) is proportional to the injected current $I^{%
\text{in}}$. In other words, we can control the polarization of the nanodisk
spin by using the spin-polarized current. We note that this is possible
since the nanodisk is a quasi-ferromagnet. Indeed, there exist no effective
magnetic field (\ref{EffecB}) in the conventional ferromagnet.

\section{Spintronic Devices and Applications}

\label{SecApplication}

We summarize the spin properties of a nanodisk and an incoming electric
current. First of all, being a quasi-ferromagnet, the nanodisk has a
definite polarization. With respect to the incoming electric current there
are three cases. (1) The polarized current, where all electrons have a
definite polarization, rotates the nanodisk spin to that of the incoming
current, as we have shown in Sec.\ref{SecReaction}. (2) The unpolarized
current, where the polarization of each electron is completely random, does
not induce any effective magnetic field. Hence it is filtered so that the
outgoing current is polarized to that of the nanodisk. (3) The partially
polarized current, where the polarization of each electron is at random but
the averaged polarization has a definite direction, induces a net effective
magnetic field. Hence it rotates the nanodisk spin to that of the incoming
current, and then is filtered so that the outgoing current is completely
polarized to the averaged polarization of the incoming current. Furthermore,
it is possible to control the nanodisk spin externally by applying magnetic
field. Then the outgoing current has the same polarization as that of the
nanodisk, irrespective of the type of incoming current. Using these
properties we propose some applications of graphene nanodisks for spintronic
devices.

\subsection{Spin Memory}

The first example is a spin memory.\cite{Recher} For a good memory device
three conditions are necessary: (i) It keeps a long life time information;
(ii) Information stored in the memory can be read out without changing the
information stored; (iii) It is possible to change the information
arbitrarily.

First, since the life time of the nanodisk quasi-ferromagnet is very long
compared to the size\cite{EzawaDisk},%
\begin{equation}
\tau _{\text{ferro}}\propto N^{2},
\end{equation}%
we may use the nanodisk spin as an information. Next, we can read-out this
information by applying a spin-unpolarized current. The outgoing current
from a nanodisk is spin-polarized to the direction of the nanodisk spin.
Thus we can obtain the information of the nanodisk spin by observing the
outgoing current. Finally, the direction of the nanodisk spin can be
controlled by applying a spin-polarized current into the nanodisk.

Thus, the nanodisk spin satisfies the conditions as a memory device. The
important point is that the size is of the order of nanometer, and it is
suitable as a nanodevice.

\subsection{Spin Amplifier}

\begin{figure}[h]
\includegraphics[width=0.35\textwidth]{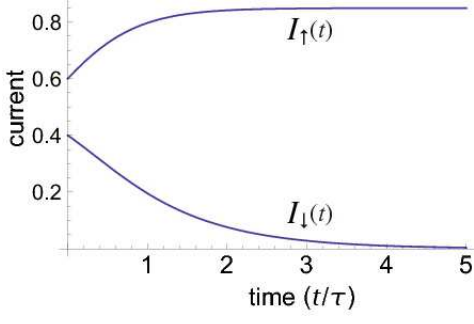}
\caption{Relaxation process of spin amplifier. The horizontal axis is the
time $t/\protect\tau $ and the vertical axis is the current $I_{\uparrow }$
and $I_{\downarrow }$. We have set $I_{\uparrow }^{\text{in}}=0.6\protect%
\sqrt{2}$, $I_{\downarrow }^{\text{in}}=0.4\protect\sqrt{2}$. The currents
saturate after enough time ($t\gtrsim 2\protect\tau $ for $I_{\uparrow }$, $%
t\gtrsim 5\protect\tau $ for $I_{\downarrow }$), and the amplification ratio
is $3$ in this example.}
\label{FigAmpRelax}
\end{figure}

The second example is a spin amplifier. We inject a partially-polarized-spin
current, whose average direction we take to be up for definiteness. Thus, $%
I_{\uparrow }^{\text{in}}>I_{\downarrow }^{\text{in}}>0$. On the other hand,
the direction of the nanodisk spin is arbitrary. Since spins in the nanodisk
feel an effective magnetic field proportional to $I_{\uparrow }^{\text{in}%
}-I_{\downarrow }^{\text{in}}$, they are forced to align with that of the
partially-polarized-spin current after making damped precession. By using (%
\ref{angle}) the time dependence is given by \beginABC%
\begin{align}
I_{\uparrow }\left( t\right) & =I_{\uparrow }^{\text{in}}\cos \frac{\theta
\left( t\right) }{2}=I_{\uparrow }^{\text{in}}\frac{1}{\sqrt{1+\exp \left[
-2t/\tau _{\text{filter}}\right] }}, \\
I_{\downarrow }\left( t\right) & =I_{\downarrow }^{\text{in}}\sin \frac{%
\theta \left( t\right) }{2}=I_{\downarrow }^{\text{in}}\frac{1}{\sqrt{1+\exp %
\left[ 2t/\tau _{\text{filter}}\right] }},
\end{align}%
\endABC where we have set $t_{0}=0$, which means $\theta \left( 0\right)
=\pi /2$: See Fig.\ref{FigAmpRelax}. The outgoing current is initially given
by%
\begin{equation}
I_{\uparrow }\left( 0\right) =\frac{1}{\sqrt{2}}I_{\uparrow }^{\text{in}%
},\qquad I_{\downarrow }\left( 0\right) =\frac{1}{\sqrt{2}}I_{\downarrow }^{%
\text{in}}.
\end{equation}%
as a function of the incoming current $I_{\sigma }^{\text{in}}$. The time
scale is given by the relaxation time (\ref{RelaxTime}).

After enough time $t-t_{0}\gg \tau $, all spins in the nanodisk take the up
direction and hence the outgoing current $I_{\sigma }^{\text{out}}\equiv
\lim_{t\rightarrow \infty }I_{\sigma }\left( t\right) $ is the perfectly
up-polarized one, 
\begin{equation}
I_{\uparrow }^{\text{out}}=I_{\uparrow }^{\text{in}},\qquad I_{\downarrow }^{%
\text{out}}=0.
\end{equation}%
Consequently, the small difference $I_{\uparrow }^{\text{in}}-I_{\downarrow
}^{\text{in}}$\ is amplified to the large current $I_{\uparrow }^{\text{in}}$%
. The amplification ratio is given by 
\hbox{$I_{\uparrow
}^{\text{in}}/(I_{\uparrow }^{\text{in}}-I_{\downarrow }^{\text{in}})$},
which can be very large. This effect is very important because the signal of
spin will easily suffer from damping by disturbing noise in leads. By
amplifying the signal we can make circuits which are strong against noises.

\subsection{Spin Valve and Spin-Field-Effect Transistor}

\begin{figure}[h]
\includegraphics[width=0.37\textwidth]{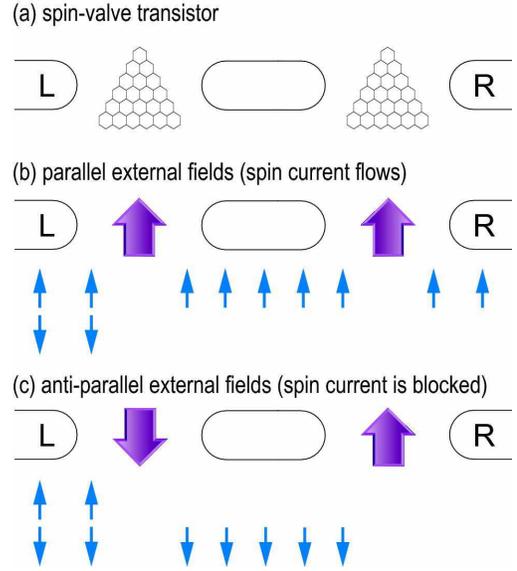}
\caption{Illustration of spin valve with unpolarized current incoming from
left. Spins of each nanodisk are controled by the external magnetic field.
The unpolarized current is filtered by the left nanodisk, and only electrons
whose spin direction is the same as the nanodisk spin go through the
centeral lead. (a) The spin valve is made of two nanodisks with the same
size, which are connected with leads. (b) We apply the external magnetic
field in the same direction. The spin-polarized outgoing current flows. (c)
We apply the external magnetic field in the opposite directions. The
outgoing current is blocked.}
\label{FigSpinValve}
\end{figure}

The third example is a spin valve, or giant magnetoresistance effect [Fig.%
\ref{FigSpinValve}].\cite{Fert,Grunberg,OhnoValve} We set up a system
composed of two nanodisks sequentially connected with leads. We apply
external magnetic field, and control the spin direction of the first
nanodisk to be 
\begin{equation}
\left\vert \theta \right\rangle =\cos \frac{\theta }{2}\left\vert \uparrow
\right\rangle +\sin \frac{\theta }{2}\left\vert \downarrow \right\rangle ,
\label{SpinValveA}
\end{equation}%
and that of the second nanodisk to be 
\begin{equation}
\left\vert 0\right\rangle =\left\vert \uparrow \right\rangle .
\end{equation}%
We inject an unpolarized-spin current to the first nanodisk. The spin of the
lead between the two nanodisks is polarized into the direction of $%
\left\vert \theta \right\rangle $. Subsequently the current is filtered to
the up-spin one by the second nanodisk. The outgoing current from the second
nanodisk is%
\begin{equation}
I_{\uparrow }^{\text{out}}=I_{\uparrow }\cos \frac{\theta }{2}.
\end{equation}%
Since we can arrange the angle $\theta $ externally, we can control the
magnitude of the up-polarized current from zero to one. In this sense the
system act as a spin valve.

The forth example is a spin-field-effect transistor\cite{Datta} [Fig.\ref%
{FigSpinTrans}]. We again set up a system composed of two nanodisks
sequentially connected with leads. We now apply the same external magnetic
field to both these nanodisks, and fix their spin direction to be up, 
\begin{equation}
\left\vert 0\right\rangle =\left\vert \uparrow \right\rangle .
\end{equation}%
As an additional setting, we use a lead between the two nanodisks possessing
a strong Rashba-type spin-orbit coupling\cite{Rashba}, 
\begin{equation}
H_{\text{R}}=\frac{\lambda }{\hbar }\left( p_{x}\sigma ^{y}-p_{y}\sigma
^{x}\right) .
\end{equation}%
Spins make precession while they pass through the lead. The spin-rotation
angle is given by\cite{Zutic}%
\begin{equation}
\Delta \theta =2\lambda m^{\ast }L/\hbar ,  \label{SpinValveTransA}
\end{equation}%
where $m^{\ast }$ is the electron effective mass in the lead and $L$ is the
length of the lead. We can control $\Delta \theta $\ by changing the
coupling strength $\lambda $ externally by applying electric field.\cite%
{NittaG} The outgoing current from the second nanodisk is%
\begin{equation}
I_{\uparrow }^{\text{out}}=I_{\uparrow }\cos \frac{\Delta \theta }{2}.
\end{equation}%
Since we can arrange the angle $\Delta \theta $ by applying electric field
and control the magnitude of the up-polarized current from zero to one, we
expect the system acts as a spin-field-effect transistor as in the
conventional case.

\begin{figure}[h]
\includegraphics[width=0.37\textwidth]{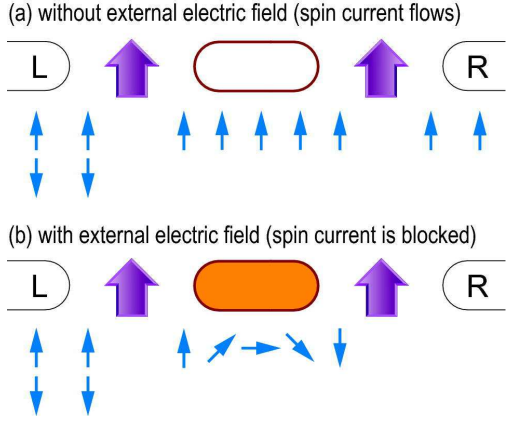}
\caption{Illustration of spin-field-effect transistor with unpolarized
current coming from left. We apply the same external magnetic field to both
of the nanodisks. The unpolarized current is filtered by the left nanodisk,
and only up-spin electrons go through the centeral lead. (a) Without
external electric field, the electron spin in the central lead does not
rotate, and the outgoing spin-polarized current exists. (b) With an
appropriate external electric field, since the electron spin in the central
lead rotates by the Rashba-type interaction, the outgoing current does not
exist. }
\label{FigSpinTrans}
\end{figure}

\subsection{Spin Diode}

\begin{figure}[h]
\includegraphics[width=0.43\textwidth]{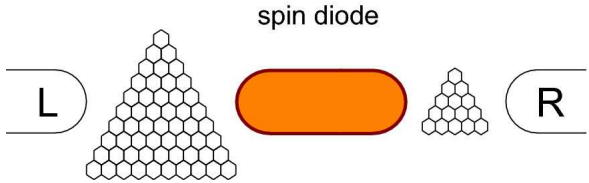}
\caption{Illustration of spin diode made of two nanodisks with different
size. By controlling the bias voltage $\Delta \protect\mu $, the current
flows from the left lead to the right lead ($\Delta \protect\mu >0$), or in
the opposite way ($\Delta \protect\mu <0$). The incoming current is
unpolarized, which is made polarized by the first nanodisk. The electron
spin in the central lead is rotated by the Rashba-type interaction.}
\label{FigSpinDiode}
\end{figure}

The fifth example is a spin diode [Fig.\ref{FigSpinDiode}]. We use a system
similar to the spin-field-effect transistor but with the following
differences. First, two nanodisks have different sizes. When the left
nanodisk is larger than the right nanodisk, the relaxation time of the left
nanodisk $\tau _{\text{L}}(\equiv \tau _{\text{filter}}^{\text{L}})$ is
larger than that of the right nanodisk $\tau _{\text{R}}(\equiv \tau _{\text{%
filter}}^{\text{R}})$,%
\begin{equation}
\tau _{\text{L}}>\tau _{\text{R}}.  \label{DiodeRelax}
\end{equation}%
Second, the applied magnetic field is taken so small that the nanodisk spin
can be controlled by a polarized current. For definiteness we take the
direction of the magnetic field to be up. Third, the lead has the
Rashba-type interaction such that the rotation angle is $\Delta \theta =\pi
-\delta $ with small $\delta $, say, $\delta \simeq 0.1\pi $. When no
currents enter the nanodisk, the direction of two nanodisk spins is
identical due to a tiny external magnetic field. When we inject the current
in this state, the net outgoing current is very small,%
\begin{equation}
I^{\text{out}}=\cos \frac{\pi -\delta }{2}\simeq 0.
\end{equation}%
This is the "off" state of spin diode.

Let us inject an unpolarized pulse square current to the system, starting at 
$t=t_{i}$ and finishing at $t=t_{f},$%
\begin{equation}
I_{\sigma }\left( t\right) =I^{\text{in}}\theta (t-t_{i})\theta (t_{f}-t),
\end{equation}%
where $\sigma $ denotes the spin. The system become the "on" state by the
pulse. When the bias voltage is such that $\Delta \mu >0$, the current flows
into the left nanodisk and then into the right nanodisk. The left nanodisk
acts as a spin filter. The current in the central lead is initially
up-polarized but is rotated by the angle $\Delta \theta $ due to the
Rashba-type coupling effect. Then it enters the right nanodisk. Since the
relaxation time is $\tau _{\text{R}}$, the total spin-dependent charge
carried by the current is given by\beginABC%
\begin{align}
Q_{\uparrow }=& \int_{t_{i}}^{t_{f}}I_{\uparrow }dt=Q_{\uparrow }\left(
t_{f}\right) -Q_{\uparrow }\left( t_{i}\right) , \\
Q_{\downarrow }=& \int_{t_{i}}^{t_{f}}I_{\downarrow }dt=Q_{\downarrow
}\left( t_{f}\right) -Q_{\downarrow }\left( t_{i}\right) 
\end{align}%
\endABC with\beginABC%
\begin{align}
Q_{\uparrow }\left( t\right) =& I^{\text{in}}\tau _{\text{R}}\sinh ^{-1}%
\left[ \exp \left[ \left( t-t_{0}\right) /\tau _{\text{R}}\right] \right] ,
\\
Q_{\downarrow }\left( t\right) =& I^{\text{in}}\tau _{\text{R}}\sinh ^{-1}%
\left[ \exp \left[ -\left( t-t_{0}\right) /\tau _{\text{R}}\right] \right] .
\end{align}%
\endABC On the other hand, when $\Delta \mu <0$, the current enter the right
nanodisk and goes out from the left nanodisk. Since the relaxation time is $%
\tau _{\text{L}}$, the total spin-dependent charge is given by the above
formulas but with the replacement of $\tau _{\text{R}}$ by $\tau _{\text{L}}$%
. Because the size of two nanodisk are different, these two currents behave
in a different way.

\begin{figure}[h]
\includegraphics[width=0.43\textwidth]{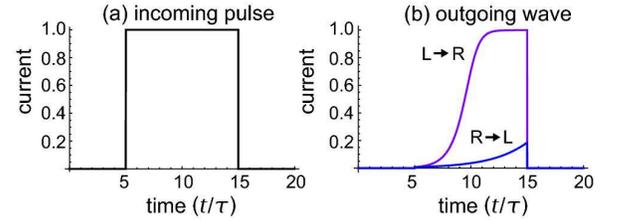}
\caption{Pulse wave. The horizontal axis is the time and the vertical axis
is the current. (a) The incoming pulse wave. (b) The outgoing waves for $%
L\rightarrow R$ and $R\rightarrow L$. They are very different for the same
incoming pulse wave due to the difference in the nanodisk size. We have set $%
t_{i}=5,t_{f}=15$, and $\protect\tau _{\text{L}}=3\protect\tau _{\text{R}}$.}
\label{FigPulse}
\end{figure}

We define the direction dependent total charge $Q_{\sigma }\left(
L\rightarrow R\right) $ and $Q_{\sigma }\left( R\rightarrow L\right) $,
where $Q_{\sigma }\left( L\rightarrow R\right) $ is the total charge with
the spin $\sigma $ when charges flow from left to right, while $Q_{\sigma
}\left( R\rightarrow L\right) $ is the total charge when charges flow from
right to left. We find the relation%
\begin{equation}
Q_{\uparrow }\left( L\rightarrow R\right) >Q_{\uparrow }\left( R\rightarrow
L\right) \gg Q_{\downarrow }\left( R\rightarrow L\right) >Q_{\downarrow
}\left( L\rightarrow R\right)
\end{equation}%
from (\ref{DiodeRelax}), which implies the up (down) component increases
(decreases) from the initial value.

\begin{figure}[h]
\includegraphics[width=0.4\textwidth]{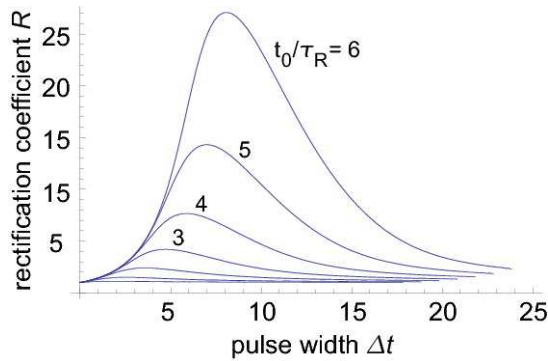}
\caption{The vertical axis is the rectification coefficient $R$. The
horizontal axis is the pulse width $\Delta t=t_{f}-t_{i}$. We plot $R$ for
various initial values $t_{0}/\protect\tau _{\text{R}}=\log \tan \left[
\left( \protect\pi -\protect\delta \right) /2\right] $. We have set $\protect%
\tau _{\text{L}}=3\protect\tau _{\text{R}}$.}
\label{FigPulsePeak}
\end{figure}

We define the rectification coefficient by%
\begin{equation}
R=\frac{Q_{\uparrow }\left( L\rightarrow R\right) +Q_{\downarrow }\left(
L\rightarrow R\right) }{Q_{\uparrow }\left( R\rightarrow L\right)
+Q_{\downarrow }\left( R\rightarrow L\right) },
\end{equation}%
which is approximately equals to 
\begin{equation}
R\simeq \frac{Q_{\uparrow }\left( L\rightarrow R\right) }{Q_{\uparrow
}\left( R\rightarrow L\right) }.
\end{equation}%
We illustrate this rectification coefficient as a function of the pulse
width $\Delta t=t_{f}-t_{i}$ for various initial phases $\theta \left(
0\right) $ or equivalently $t_{0}$ in Fig.\ref{FigPulsePeak}. Each curve has
a peak structure. Hence, when the relaxation-time ratio $\tau _{\text{L}%
}/\tau _{\text{R}}$ and the initial phase $\theta \left( 0\right) $ are
given, it is possible to optimize the width $\Delta t$ so that the
rectification coefficient $R$ is maximized. This maximized value of $R$
diverges as $\theta \left( 0\right) \rightarrow \pi $. However, the
relaxation time diverges as well. It would be efficient to take the initial
phase $\theta \left( 0\right) $ to make $t_{0}/\tau _{\text{R}}\simeq 5$ for
a spin diode.

\section{Conclusions}

\label{SecConclusion}

We have studied the electromagnetic properties of the zigzag trigonal
nanodisk by projecting the system to the zero-energy sector. We may regard
it as a quasi-ferromagnet characterized by the exchange energy as large as
the Coulomb energy. The system is well approximated by the SU(N)
Heisenberg-Hubbard model. The relaxation time is finite but quite large even
if the size is very small. Being a ferromagnet, it can be used as a spin
filter. Namely, only electrons with spin parallel to the spin of the
nanodisk can go through it. Additionally, it has a novel feature that it is
not a rigid ferromagnet. The incoming spin-polarized current can rotate the
nanodisk spin itself. Combining the advantages of both these properties, we
have proposed a rich variety of spintronic devices, such as spin memory,
spin amplifier, spin valve, spin-field-effect transistor and spin diode.
Graphene nanodisks could well be basic components of future nanoelectronic
and spintronic devices.

I am very much grateful to N. Nagaosa and S. Tarucha for many fruitful
discussions on the subject.

\end{document}